\documentclass[epj,nopacs,fleqn]{svjour}
\usepackage{epsfig,amsmath,amssymb}
\usepackage{graphicx}
\usepackage{slashed}

\newcommand{\nn}{\nonumber}

\newcommand{\lam}{\lambda}
\newcommand{\sig}{\sigma}
\newcommand{\M}{{\cal M}}
\newcommand{\Mpl}{\overline{M}_{\rm Pl}}
\newcommand{\gld}{\tilde G}
\newcommand{\no}{\tilde{\chi}^0_1}
\newcommand{\se}{\tilde e}
\newcommand{\go}{\tilde g}
\newcommand{\sq}{\tilde q}

\begin{document}

\title{Associated production of light gravitinos 
       in $e^+e^-$ and $e^-\gamma$ collisions}

\author{
 K.~Mawatari\inst{1}\fnmsep\thanks{e-mail: 
            kentarou.mawatari@vub.ac.be},
 B.~Oexl\inst{1}\fnmsep\thanks{e-mail: bettina.oexl@vub.ac.be},  
 Y.~Takaesu\inst{2}\fnmsep\thanks{e-mail: takaesu@post.kek.jp} }
\institute{
 Theoretische Natuurkunde and IIHE/ELEM, Vrije Universiteit Brussel,\\
 and International Solvay Institutes,
 Pleinlaan 2, B-1050 Brussels, Belgium
 \and
 KEK Theory Center, and Sokendai, Tsukuba 305-0801, Japan
 }

\abstract{
Light gravitino productions in association with a neutralino
(selectron) in $e^+e^-$ ($e^-\gamma$) collisions are restudied in a
scenario that the lightest supersymmetric particle is a gravitino
and the produced neutralino (selectron)  promptly decays into a photon
(electron) and a gravitino.  
We explicitly give the helicity amplitudes for the production processes 
by using the effective goldstino interaction Lagrangian, and present the
cross sections with different collision energies and mass spectra.
We also examine selection efficiencies by kinematical cuts and beam
polarizations for the signal and background processes, and show that the 
energy and angular distributions of the photon (electron) can explore
the mass of the $t$-channel exchange particle as well as the mass of
the decaying particle at a future $e^+e^-$ ($e^-\gamma$) collider. 
}

\titlerunning{Associated production of light gravitinos 
              in $e^+e^-$ and $e^-\gamma$ collisions}
\authorrunning{K.~Mawatari, B.~Oexl, Y.~Takaesu}

\maketitle

\vspace*{-10.5cm}
\noindent KEK-TH-1462
\vspace*{8.83cm}


\section{Introduction}\label{intro}

Gravitinos are spin-3/2 superpartners of gravitons in local 
supersymmetric extensions to the Standard Model (SM). 
Since the gravitino becomes massive via the super-Higgs mechanism, 
its mass is related to the scale of supersymmetry (SUSY) breaking as
well as the Planck scale like 
\begin{align}
 m_{3/2}\sim (M_{\rm SUSY})^2/M_{\rm Pl}.
\end{align}
This implies that the gravitino can take a wide range of mass,
depending on the SUSY breaking scale, from eV up to scales beyond
TeV, and provide rich phenomenology in particle physics
as well as in cosmology~\cite{Giudice:1998bp}.  
While the interactions of the helicity $\pm3/2$ components of the
gravitino are suppressed by the Planck scale, those of the helicity
$\pm1/2$ components are suppressed by the SUSY breaking scale 
if the gravitino mass is much smaller than the energy scale of the
interactions, due to the goldstino equivalence theorem, 
and can be important even for collider phenomenology.

Gravitino productions in association with a SUSY particle are known
processes which become significant at colliders when the mass of
the gravitino is very light as $m_{3/2}\sim\cal O$($10^{-2}$ eV) or
less, since the cross sections are inversely proportional to the square
of the gravitino mass 
\begin{align}
 \sigma\propto 1/m_{3/2}^2.
\end{align}
Such a very light gravitino is suggested by the context of no-scale
supergravity~\cite{Ellis:1984kd,Lopez:1992ni} and some extra-dimensional
models~\cite{Gherghetta:2000qt}, while typical gauge-mediated
SUSY breaking (GMSB) scenarios expect a mass of 1 eV--10
keV~\cite{Giudice:1998bp}. 
Several studies on the associated gravitino productions have been
performed so far,
for instance, $\no$-$\gld$ productions in
$e^+e^-$~\cite{Fayet:1986zc,Dicus:1990vm,Lopez:1996gd,Lopez:1996ey}
and hadronic~\cite{Lopez:1996ey} collisions, $\se$-$\gld$ productions 
in $e\gamma$ collisions~\cite{Gopalakrishna:2001cm}, and 
$\go$-$\gld$~\cite{Dicus:1989gg,Kim:1997iwa,Klasen:2006kb} and
$\sq$-$\gld$~\cite{Kim:1997iwa,Klasen:2006kb} productions in hadronic
collisions. 
When the associated SUSY particle is the next-to-lightest supersymmetric
particle (NLSP) and promptly decays into a SM particle and a LSP
gravitino,
the above production processes lead to particular collider signatures,
such as $\gamma+\slashed E$, $e+\slashed E$, and jet$+\slashed E$, where
the missing energy is carried away by two gravitinos, and these signals
set mass bounds on the gravitino and the other SUSY particles.
The current experimental bound on the gravitino mass
from the single-photon plus missing-energy signal%
\footnote{We note that a two-photon plus missing-energy signal, where
the two photons come from two neutralino decays, does not provide any
constraint on the gravitino mass.}  
in $\no$-$\gld$ associated productions is given by the LEP experiment as
a function of the neutralino and selectron
masses~\cite{Abdallah:2003np}, e.g. 
\begin{align}
 m_{3/2}\gtrsim 10^{-5}\ {\rm eV}
\end{align}
for $m_{\no}=140$ GeV and $m_{\se}=150$ GeV.
We note that the Tevatron also set a similar bound on the gravitino mass
for the $\gamma+\slashed E$~\cite{Acosta:2002eq} and 
jet$+\slashed E$~\cite{Affolder:2000ef} channels, where it is assumed,
however, that all SUSY particles except the gravitino are too heavy to
be produced on-shell~\cite{Brignole:1998me}.

While the previous searches for gravitino productions have been somewhat
restricted due to limitations of simulation tools, in the recent
paper~\cite{Hagiwara:2010pi} implementation of the spin-3/2 gravitino in
{\tt MadGraph/MadEvent}  
{\tt (MG/ME)}~\cite{Stelzer:1994ta,Maltoni:2002qb,Alwall:2007st} was
reported, where new {\tt HELAS} 
({\tt HEL}icity {\tt A}mplitude {\tt S}ubroutines)~\cite{Hagiwara:1990dw}
codes were introduced to calculate helicity amplitudes with massive
spin-3/2 gravitinos. 
They are implemented in such a way that amplitudes with external
gravitinos can be generated automatically by {\tt MG/ME}.%
\footnote{The spin-3/2 functionality is available in 
 {\tt MG/ME V4.5}~\cite{Alwall:2007st}.}
Since goldstinos appear as the longitudinal modes of massive gravitinos
and their interactions become dominant over the transverse modes in
high-energy processes, two of the authors also implemented effective
goldstino interactions~\cite{Mawatari:2011jy} as an alternative to the
gravitino code. 

In this paper, we revisit the following two processes by using the
gravitino implemented {\tt MG/ME} mentioned above.
First, we study associated gravitino productions with a neutralino which
promptly decays into a photon and a gravitino in $e^+e^-$ collisions,
\begin{align}
 e^+e^-\to\no\gld\to\gamma\gld\gld, \nn
\end{align}
in the context of a neutralino NLSP with a gravitino LSP.
In order to investigate the production cross section and distributions
of the photon in detail, we explicitly give the helicity amplitudes for
the production process by using the effective goldstino interaction
Lagrangian, and present the cross sections with different mass spectra
and different energies especially for a future linear collider. 
We also examine selection efficiencies by kinematical cuts and beam
polarizations for the signal and SM background processes, and show that
the energy and angular distributions of the photon coming from the
neutralino decay can explore the mass of the $t$-channel exchange
selectrons as well as the mass of the decaying neutralino. 

Second, we consider gravitino productions in association with a
selectron which subsequently decays into an electron and a gravitino at
an $e\gamma$ collider, which is an option at a future linear
collider~\cite{:2007sg},  
\begin{align}
 e^-\gamma\to\se^-\gld\to e^-\gld\gld, \nn
\end{align}
in a slepton co-NLSP scenario with a gravitino LSP.
We present the explicit helicity amplitudes for the production
process, and discuss the mono-electron plus missing-energy signal,
including the Compton back-scattered photon energy
spectrum~\cite{Ginzburg:1981ik,Ginzburg:1982yr} for incident photons. 
While the heavy-mass limit for all SUSY particles except gravitino and
selectron are assumed in Ref.~\cite{Gopalakrishna:2001cm}, we take into
account the $t$-channel intermediate neutralinos and show a
possibility to determine their mass in the signal distributions. 

We note in passing that all the helicity amplitudes we present are
easily applicable to $q\bar q\to\go\gld$ and to $qg\to\sq\gld$
subprocesses for hadron colliders. 

The paper is organized as follows:
In Sect.~\ref{sec:n1nlsp} neutralino-gravitino productions in
electron-positron collisions are considered, and in
Sect.~\ref{sec:senlsp} selectron-gravitino productions in
electron-photon collisions are studied.  
Sect.~\ref{sec:summary} is devoted to our summary.
In Appendix~\ref{sec:lagrangian} we give the effective goldstino
interaction Lagrangian relevant to our study, and in
Appendix~\ref{sec:n1decay} we briefly mention neutralino decays into a
photon and a gravitino.

\section{Neutralino-gravitino production in $e^+e^-$ collisions}
\label{sec:n1nlsp}

In this section, we consider a scenario of a neutralino NLSP with a
gravitino LSP, and study associated gravitino productions with a
neutralino which promptly decays into a photon and a gravitino in
$e^+e^-$ collisions,  
\begin{align}
 e^+e^-\to\no\gld\to\gamma\gld\gld,
\label{process1}
\end{align}
leading to a mono-photon plus missing-energy signal. 

\begin{figure}[b]
 \epsfig{file=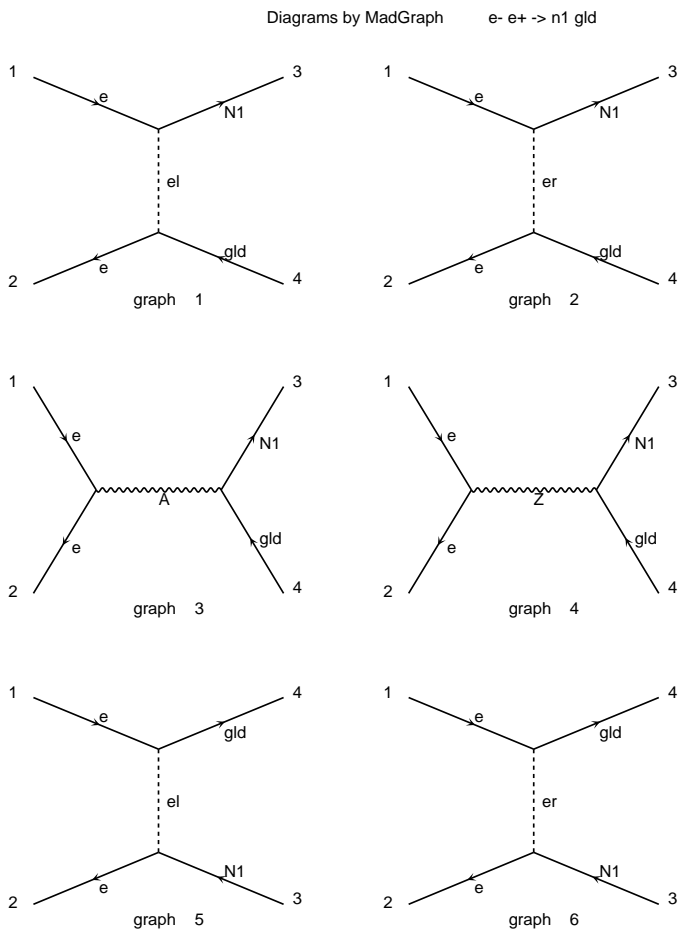,width=1\columnwidth,clip}
 \caption
 {Feynman diagrams for the $\no$-$\gld$ production in $e^+e^-$
 collisions, generated by {\tt MadGraph}~\cite{Mawatari:2011jy}.  
 {\tt N1}, {\tt gld}, {\tt el}, and {\tt er} denote a lightest
 neutralino, a gravitino, a left-handed selectron, and a right-handed
 selectron, respectively.}
\label{fig:diagram_n1}
\end{figure} 

\begin{table*}
\centering
\begin{tabular}{cc|rccccc}
 $\lam$ & $\lam_3\lam_4$ 
  &&& $\hat{\M}^{s}$ & $\hat{\M}^{t}$ & $\hat{\M}^{u}$ & \\ \hline
 $\pm$ & $\pm\mp$ & $(1+\cos{\theta})$ & \big[ & 
  $\frac{m_{\tilde{\chi}}^2}{s}C^s_{\pm}$ && 
  $-\frac{m^2_{\se_{\pm}}}{u-m^2_{\se_{\pm}}}C^{\se\tilde\chi_1}_{\pm}$
  & \big]\\
 $\pm$ & $\mp\pm$ & $-(1-\cos{\theta})$ & \big[ &
  $\frac{m_{\tilde{\chi}}^2}{s}C^s_{\pm}$ & 
  $-\frac{m^2_{\se_{\pm}}}{t-m^2_{\se_{\pm}}}C^{\se\tilde\chi_1}_{\pm}$ 
  && \big]\\
 $\pm$ & $\pm\pm$ & 
  $\pm\frac{m_{\tilde{\chi}}}{\sqrt{s}}\sin{\theta}$ & \big[ &
  $C^s_{\pm}$ & 
  $-\frac{m^2_{\se_{\pm}}}{t-m^2_{\se_{\pm}}}C^{\se\tilde\chi_1}_{\pm}$ 
  && \big]\\
 $\pm$ & $\mp\mp$ & 
  $\mp\frac{m_{\tilde{\chi}}}{\sqrt{s}}\sin{\theta}$ & \big[ & 
  $C^s_{\pm}$ && 
  $-\frac{m^2_{\se_{\pm}}}{u-m^2_{\se_{\pm}}}C^{\se\tilde\chi_1}_{\pm}$ 
  & \big] 
\end{tabular}
\caption{The reduced helicity amplitudes $\hat\M_{\lam,\lam_3\lam_4}$
 for $e^-_{\lam}e^+_{-\lam}\to {\no}{}^{}_{\lam_3}\gld^{}_{\lam_4}$.} 
\label{table:n1}
\end{table*}

\subsection{Helicity amplitudes}

Here we present the helicity amplitudes explicitly for the production
process: 
\begin{align}
 e^-\Big(p_1,\frac{\lam_1}{2}\Big)+e^+\Big(p_2,\frac{\lam_2}{2}\Big)\to
 \no\Big(p_3,\frac{\lam_3}{2}\Big)+\gld\Big(p_4,\frac{\lam_4}{2}\Big),
\label{process_n1}
\end{align}
where the four-momentum ($p_i$) and helicity ($\lam_i=\pm 1$) of each
particle are defined in the center-of-mass (CM) frame of the $e^+e^-$
collisions. 
Throughout our study only the helicity $\pm1/2$ components
of the gravitino, i.e. goldstinos, are considered.  
In the massless limit of $e^{\pm}$, one can find that all the
amplitudes are zero when both the electron and the positron have the
same helicity, or $\lam_1=\lam_2$. 
In addition, for the $\lam_1=+1$ ($\lam_1=-1$) case, only the
right-handed (left-handed) selectron can contribute to the total
amplitudes. Therefore, the helicity amplitudes for the above process can
be expressed as the sum of $s$-, $t$-, and $u$-channel amplitudes:
\begin{align}
 \M^{}_{\lam,\lam_3\lam_4}=\M^{s}_{\lam,\lam_3\lam_4}
  +\M^{t}_{\lam,\lam_3\lam_4}+\M^{u}_{\lam,\lam_3\lam_4}
\label{amp_n1gld}
\end{align}
with $\lam\equiv\lam_1=-\lam_2$, where each amplitude with $\lam=+1$
($\lam=-1$) corresponds to the Feynman graph 3+4, 2 (1), and 6 (5),
respectively, in Fig.~\ref{fig:diagram_n1}.

We first present the amplitudes based on the effective
goldstino interaction Lagrangian, given in
Appendix~\ref{sec:lagrangian}, in the usual four-spinor basis:
\begin{subequations}
\begin{align}
 i{\M}^s_{\lam,\lam_3\lam_4}
  &=\frac{e\,C^s_{\lam}\,m_{\no}}{2\sqrt{6}\,\Mpl m_{3/2}}\frac{1}{s}\,
    \bar{v}(p_2,-\lam)\gamma^{\mu}u(p_1,\lam) \nn\\
   &\ \times\bar{u}(p_3,\lambda_3)
         [\slashed{p}_3+\slashed{p}_4,\gamma_{\mu}]v(p_4,\lambda_4), 
  \label{zamp}\\
 i{\M}^{t}_{\lam,\lam_3\lam_4}
  &=\frac{-\sqrt{2}\,e\,{C^{\se\tilde\chi_1}_{\lam}}m_{\se_{\lam}}^2}
             {\sqrt{3}\,\Mpl m_{3/2}}
        \frac{1}{t-m^2_{\tilde{e}_{\lam}}} \nn\\
   &\ \times\bar{u}(p_3,\lambda_3)u(p_1,\lam)\,
    \bar{v}(p_2,-\lam)v(p_4,\lambda_4), \\
 i{\M}^{u}_{\lam,\lam_3\lam_4}
  &=\frac{-\sqrt{2}\,e\,C^{\se\tilde\chi_1}_{\lam}m_{\se_{\lam}}^2}
             {\sqrt{3}\,\Mpl m_{3/2}}
    \frac{1}{u-m^2_{\tilde{e}_{\lam}}} \nn\\
   &\ \times\bar{u}(p_4,\lambda_4)u(p_1,\lam)\,
               \bar{v}(p_2,-\lam)v(p_3,\lambda_3),
\end{align}
\end{subequations}
where $\Mpl\equiv M_{\rm Pl}/\sqrt{8\pi}\sim2.4\times10^{18}$ GeV is the
reduced Planck mass, $m_{\se_{\pm}}$ denotes the right-/left-handed
selectron mass for notational convenience, and  
\begin{align}
 C^s_{\lam}= C^{\gamma\tilde\chi_1}
            -\frac{s}{s-m^2_Z+im_Z\Gamma_Z}g_{\lam}C^{Z\tilde\chi_1}
\end{align}
with $Z$-boson couplings to right- and left-handed charged leptons,
\begin{align}
 g_+=\frac{\sin\theta_W}{\cos\theta_W}\quad {\rm and}\quad
 g_-=\frac{-1+2\sin^2\theta_W}{2\sin\theta_W\cos\theta_W},
\label{zcouplings}
\end{align}
respectively.%
\footnote{Strictly speaking, the $Z$-exchange amplitude in \eqref{zamp}
 is valid only for $\sqrt{s}\gg m_Z$ since massless gauge bosons are
 assumed in the effective Lagrangian~\eqref{L_int}.} 
The couplings related to the neutralino mixing defined by
$X_i=U_{ij}\tilde\chi^0_j$ in the
$X=(\tilde B,\tilde W^3,\tilde H^0_d,\tilde H^0_u)$ basis, where
$U_{ij}$ is taken to be real, are  
\begin{align}
 C^{\gamma\tilde\chi_i}&=U_{1i}\cos{\theta_W}+U_{2i}\sin{\theta_W},\nn\\
 C^{Z\tilde\chi_i}&=-U_{1i}\sin{\theta_W}+U_{2i}\cos{\theta_W}, \nn\\
 C^{\se\tilde\chi_i}_{\pm}&=T^{\se}_{\pm}\frac{U_{2i}}{\sin{\theta_W}}
                +Y^{\se}_{\pm}\frac{U_{1i}}{\cos{\theta_W}},
\label{couplings}
\end{align}
with the $SU(2)$ charge $T^{\se}_{\pm}$ and the $U(1)$ charge
$Y^{\se}_{\pm}$ for $\se_{+/-}(=\se_{R/L})$. 
Here, for simplicity, we assume the lightest neutralino as a pure
gaugino, which makes the $\gld$-$\tilde H^0_{d,u}$-$Z$ couplings
irrelevant to our study.
It should be noted that $C^s_{\lam}$ and $C^{\se\tilde\chi_1}_{\lam}$
are related with each other as
\begin{align}
 C^s_{\lam}\sim-C^{\se\tilde\chi_1}_{\lam}
               +{\cal O}\Big(\frac{m^2_Z}{s}\Big)
\label{CsCe}
\end{align}
for $\sqrt{s}\gg m_Z$; this is always the case in the following
discussions.  

To present the explicit helicity amplitudes, let us now define the
kinematical variables of the process~\eqref{process_n1} in the $e^+e^-$
laboratory frame as 
\begin{align}
 p_1^{\mu}&=\tfrac{\sqrt{s}}{2} (1,0,0,1), \nn\\
 p_2^{\mu}&=\tfrac{\sqrt{s}}{2} (1,0,0,-1), \nn\\
 p_3^{\mu}&=\tfrac{\sqrt{s}}{2} 
            \big(1+\tfrac{m^2_{\tilde{\chi}}}{s},\beta\sin\theta,0,
                 \beta\cos\theta\big), \nn\\
 p_4^{\mu}&=\tfrac{\sqrt{s}}{2}
            \big(1-\tfrac{m^2_{\tilde{\chi}}}{s},-\beta\sin\theta,0,
                 -\beta\cos\theta), 
\label{equation_momenta_1}
\end{align}
with $\beta=1-m_{\no}^2/s$. Throughout our study we neglect the
gravitino mass, except in the gravitino couplings.

For notational convenience we define the reduced helicity amplitudes,
$\hat{\M}$, as
\begin{align}
 i\M_{\lam,\lam_3\lam_4}=\frac{-e}{\sqrt{6}\,\Mpl m_{3/2}}\sqrt{\beta}\,s\, 
 \hat{\M}_{\lam,\lam_3\lam_4},
\label{ramp}
\end{align}
and these are presented in Table~\ref{table:n1}. 
The following features of the amplitudes are worth noting: 
\begin{enumerate}
\item As mentioned before, for the $\lam=+1$ ($\lam=-1$) case only
      $\se_+$ ($\se_-$) can be exchanged in the $t$- and $u$-channel
      amplitudes, and all the amplitudes are zero for
      $\lam_1=\lam_2$. \\[-2mm] 
\item The overall angular dependence is dictated by $J=1$ $d$ functions
      as 
\begin{align}
 \M_{\lam,\lam_3\lam_4}\propto 
 d^{1}_{\lam,(\lam_3-\lam_4)/2}(\theta).
\end{align}
\item $\M^s$ and $\M^{t,u}$ interfere subtractively with each other;
      especially for the $\lam_3=\lam_4$ case they almost cancel in the
      wide range of the parameter space, and hence the amplitudes with
      $\lam_3=-\lam_4$ are dominant for the most of the cases except for
      the $m_{\no}\sim m_{\se_{\pm}}$ region.  
      We note that in the very high-energy region the amplitudes with
      $\lam_3=\lam_4$ become important since $\M^s$ becomes dominant,
      making those amplitudes be proportional to $\sqrt{s}$ while the
      amplitudes with $\lam_3=-\lam_4$ are independent of $\sqrt{s}$; 
      in that region the cross section does not depend on the selectron
      masses but on the produced neutralino mass. \\[-2mm] 
\item For the $\lam_3=-\lam_4$ case, in the threshold region, where
      $t,u=-s\beta(1\mp\cos\theta)/2\to 0$, an additional $\beta$ can be
      extracted from the reduced amplitudes due to
      $C^s_{\pm}\sim-C^{\se\tilde\chi_1}_{\pm}$ in \eqref{CsCe}.  
      Together with $\beta^{1/2}$ in~\eqref{ramp}, the amplitudes are
      proportional to $\beta^{3/2}$. Therefore, including the phase
      space factor $\beta$, the threshold excitation of the total cross
      section is given by~\cite{Fayet:1986zc,Lopez:1996gd,Lopez:1996ey}%
\footnote{Note that $\beta$ is defined as $(1-m_{\no}^2/s)^{1/2}$ 
in Refs.~\cite{Lopez:1996gd,Lopez:1996ey}.} 
\begin{align}
 \sigma\propto\beta^4.
\label{beta4}
\end{align}
\item $\M^{t}$ and $\M^u$ depend on the selectron mass and become larger 
      as the selectron mass increases, while $\M^s$ is independent of
      $m_{\se}$.  
\end{enumerate}

We note that our helicity-summed amplitude squared agrees with Eq.~(28)
in~\cite{Lopez:1996ey} for the photino case, and also with Eq.~(3)
of~\cite{Klasen:2006kb} for the gluino associated process
$q\bar q\to\go\gld$ after substitutions for the masses and the couplings
as 
\begin{align}
 &m_{\no}\to m_{\go},\quad m_{\se_{R/L}}\to m_{\sq_{R/L}},\nn\\
 &e\to -g_sT^a,\quad C^s_{\lam}\to1,\quad 
  C^{\se\tilde\chi_1}_{\lam}\to-1.
\label{substitute}
\end{align}
Moreover, our analytic amplitudes are checked numerically for each
helicity combination by using the gravitino/goldstino code in 
{\tt MG/ME}~\cite{Hagiwara:2010pi,Mawatari:2011jy}.

\subsection{Cross sections and kinematical distributions}

Let us now present the total cross sections and the kinematical
distributions for the production process~\eqref{process_n1}.
The initial-helicity ($\lambda$) dependent cross section is given by
\begin{align}
 d\sigma_{\lam}=\frac{1}{2s}\frac{1}{2}\sum_{\lam_{3,4}}
                |\M_{\lam,\lam_3\lam_4}|^2d\Phi_2
\end{align}
with the two-body phase space factor $d\Phi_2$.
$\sigma_{\rm unpol}=(\sigma_++\sigma_-)/2$ is the usual spin-summed and
averaged cross section. 

\begin{figure}
 \epsfig{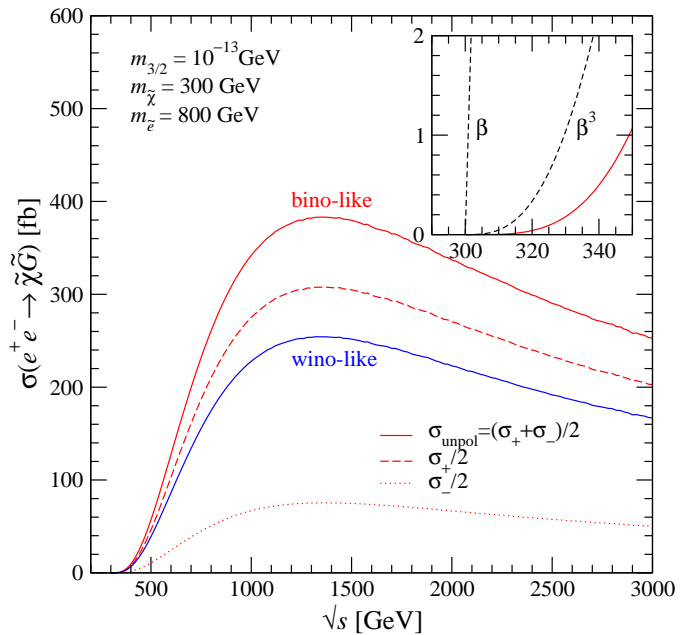}
 \caption
 {Total cross sections of associated gravitino productions with a
 bino-/wino-like neutralino in $e^+e^-$ collisions, $e^+e^-\to\no\gld$,
 for $m_{3/2}=10^{-13}$ GeV as a function of the collision energy.  
 The neutralino and selectron masses are fixed at 300 GeV and 800 GeV, 
 respectively. The initial-helicity dependent cross sections
 $\sig_{\lam}$ are shown by a dashed line for $\lam=+1$ and a dotted
 line for $\lam=-1$. The threshold region is enlarged and the
 hypothetical dependence $\sigma\propto\beta$ and $\beta^3$ with the
 same coefficient is also shown.}
\label{fig:xsec_rs_n1}
\end{figure} 

Figure~\ref{fig:xsec_rs_n1} shows total cross sections of the gravitino
productions associated with a bino-/wino-like neutralino in $e^+e^-$
collisions as a function of the CM energy $\sqrt{s}$, where the
neutralino and selectron masses are fixed as $m_{\no}=300$ GeV and
$m_{\se_+}=m_{\se_-}=800$ GeV. 
It should be stressed that the cross section scales with $m_{3/2}^{-2}$, 
and we fix the gravitino mass $m_{3/2}=10^{-13}$ GeV in our study so
that the production cross sections are around
${\cal O}(10^1\sim10^3)$~fb.  
In the figure the threshold region for the unpolarized bino-like
neutralino cross section is enlarged, and one can see that the
production cross section is strongly suppressed as shown
in~\eqref{beta4}, in contrast to the threshold excitation for the
standard fermion ($\propto\beta$) and the scalar ($\propto\beta^3$) pair 
productions~\cite{Choi:2006mr}. This is one of the particular
signatures for the associated gravitino productions.

For the case of the bino-like neutralino, or $|U_{11}|\sim1$ and
$|U_{21}|\sim0$ in~\eqref{couplings}, the cross section with
right-handed electrons ($\sig_+$) dominates the one with left-handed
($\sig_-$). For the heavy selectron case the $t$- and $u$-channel
contributions are dominant, and therefore the ratio of the
$\lam$-dependent cross sections is roughly given in terms of the
$\no$-$e$-$\se_{\pm}$ couplings as  
\begin{align}
 \frac{\sig_{\pm}}{2\sig_{\rm unpol}} \sim 
 \frac{|C^{\se\tilde\chi_1}_{\pm}|^2}
      {|C^{\se\tilde\chi_1}_+|^2+|C^{\se\tilde\chi_1}_-|^2}.
\end{align}
The bino case gives $\sig_+/2\sig_{\rm unpol}\sim 0.8$, which one can 
observe in Fig.~\ref{fig:xsec_rs_n1}.
On the other hand, for the case of the wino-like neutralino, or 
$|U_{11}|\sim 0$ and $|U_{21}|\sim 1$, the
right-handed cross section vanishes, i.e. $\sig_{\rm unpol}=\sig_-/2$.  
One can conclude that the $\no$-$\gld$ production process with polarized
electron beam can explore the neutralino mixing.
The detailed study for various neutralino mixing has been done in 
Refs.~\cite{Lopez:1996gd,Lopez:1996ey}, while we assume a bino-like
neutralino in the following analyses for simplicity, which is often the
case in GMSB; see, e.g. Fig.~4 in~\cite{Ambrosanio:1999iu}.

\begin{figure}
 \epsfig{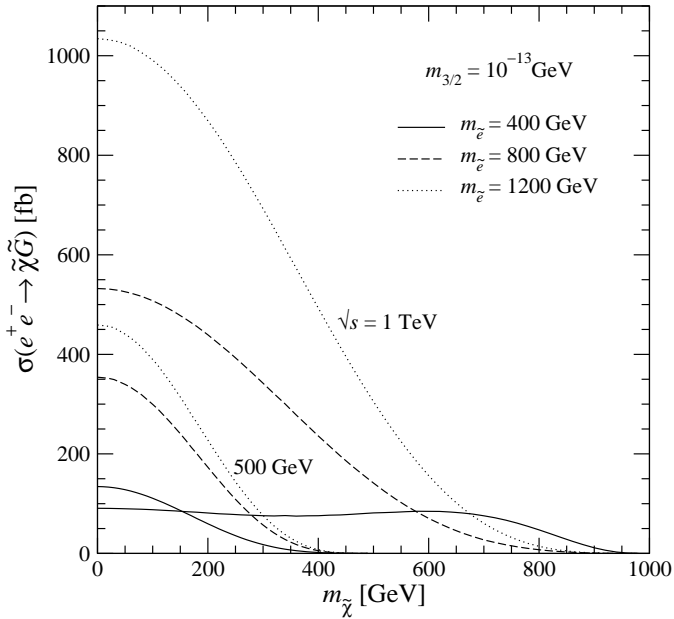}
 \caption
 {Total cross sections of $e^+e^-\to\no\gld$ at $\sqrt{s}=500$~GeV and
 1~TeV for $m_{3/2}=10^{-13}$ GeV as a function of the neutralino mass.
 The selectron masses are fixed at 400 (solid), 800 (dashed) and 1200
 (dotted) GeV, respectively.}
\label{fig:xsec_mn1}
\end{figure} 

In Fig.~\ref{fig:xsec_mn1}, the neutralino-mass dependence of the cross
sections is shown for $\sqrt{s}=500$ GeV and 1 TeV.  
Due to the threshold behavior in~\eqref{beta4}, the cross sections are
strongly suppressed as the neutralino mass is approaching the collider
energy.%
\footnote{
 For the case of $m_{\se_{\pm}}=400$ GeV at $\sqrt{s}=1$ TeV, 
 the cross section is not so strongly suppressed as $\beta^4$. This is
 because
 in this parameter region the contributions from the amplitudes with
 $\lam_3=\lam_4$ are significant and these amplitudes do not provide an
 additional suppression factor $\beta$.}
It should be emphasized here that the cross section is quite sensitive
to the mass of the $t,u$-channel intermediate selectrons, even if the
collider energy cannot reach them~\cite{Lopez:1996gd,Lopez:1996ey}. 
The heavier selectron exchange increases the cross section since the
$t,u$-channel amplitudes are proportional to the selectron mass
squared as one can see in Table~\ref{table:n1}.
We also note that, however, the goldstino couplings become too strong at some
point for heavy selectrons to perform the reliable perturbative calculations. 

\begin{figure}
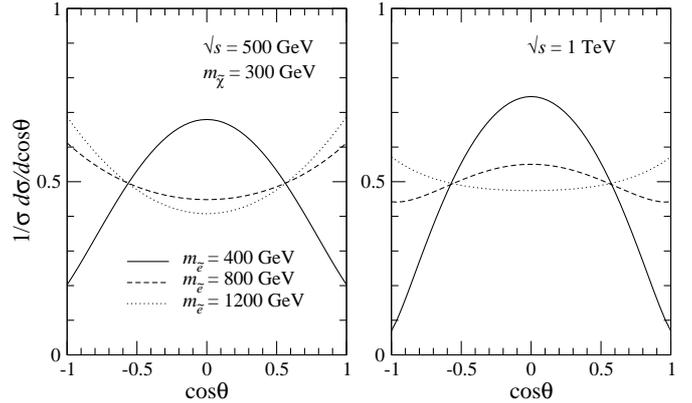

 \epsfig{file=cosn1_l.eps,height=5.3cm,clip}
 \epsfig{file=cosn1_h.eps,height=5.3cm,clip}
 \caption
 {Normalized angular distributions of the neutralino in
 $e^+e^-\to\no\gld$ at $\sqrt{s}=500$ GeV (left) and 1 TeV (right) for
 $m_{\no}=300$ GeV, where the selectron masses are taken to be 400, 800
 and 1200 GeV.}
\label{fig:cos}
\end{figure}

Before taking the neutralino decay into account, we discuss the angular 
distribution of the produced neutralino since the 
$\no\to\gamma\gld$ decay is isotropic (see Appendix~\ref{sec:n1decay}) and
hence the photon distribution is given by purely
kinematical effects of the decaying neutralino.
In Fig.~\ref{fig:cos}, the normalized $\cos\theta$ distributions of the
neutralino in $e^+e^-\to\no\gld$ at $\sqrt{s}=500$ GeV (left) and 1 TeV
(right) are shown for $m_{\no}=300$ GeV. 
One can find that not only the total cross section as shown in
Fig.~\ref{fig:xsec_mn1} but also the angular distribution is quite
sensitive to the mass of the $t,u$-channel intermediate
selectrons~\cite{Lopez:1996ey}. 
When the selectron mass is close to the neutralino mass, the cross
section is suppressed around $|\cos\theta|=1$ since a cancellation
occurs between $\M^s$ and $\M^{t,u}$ for the $\lam_3=-\lam_4$ case due
to $(t-m_{\se}^2)=-s$ for $\cos\theta=-1$ and $(u-m_{\se}^2)=-s$ for
$\cos\theta=1$; see also Table~\ref{table:n1}.
For the heavy selectron case, on the other hand, the neutralino tends to 
be produced to the forward and backward regions since the selectron
exchange diagrams are dominant and give the $(1+\cos\theta)^2$ or
$(1-\cos\theta)^2$ angular dependence. We note that the contributions
from the $\lam_3=\lam_4$ case, which could give $\sin^2\theta$
dependence, are negligible for heavy selectron masses as mentioned
before.

\begin{table*}
\centering
\begin{tabular}{lr|crrr}
 $\qquad\sigma$ [fb] & & $(P_{e^-},P_{e^+})=$ & $(0,0)$ & 
  $(0.9,0)$ & $(0.9,-0.6)$ \\
 \hline
 $\sqrt{s}=500$ GeV & $m_{\se}=400$ GeV & 
 & 15 & 23 & 37 \\
 & 800 GeV & & 48 & 75 & 119 \\
 & 1200 GeV & & 64 & 100 & 159 \\
 & SM background & & 1592 & 178 & 94 \\
 \hline
 $\sqrt{s}=1$ TeV & $m_{\se}=400$ GeV &
 & 72 & 112 & 177 \\ 
 & 800 GeV & & 320 & 494 & 785 \\
 & 1200 GeV & & 642 & 1002 & 1582 \\
 & SM background & & 1443 & 149 & 65 
\end{tabular}
 \caption
 {Cross sections in fb unit for the signal,
 $e^+e^-\to\no\gld\to\gamma\gld\gld$, and the SM
 background, $e^+e^-\to\gamma\nu\bar\nu$, at $\sqrt{s}=500$~GeV and
 1~TeV with different beam polarizations $P_{e^{\pm}}$. 
 We take  $m_{3/2}=10^{-13}$ GeV, $m_{\no}=300$ GeV, and 
 $B(\no\to\gamma\gld)=1$. 
 The minimal cuts in~\eqref{mincut} and the $Z$-peak cut in \eqref{zcut}
 are taken into account.}  
\label{table:photon}
\end{table*}

Let us now turn to the simulation for the single-photon signal with
missing energy. The partial decay rate of the neutralino decay into a
photon and a gravitino is given by (see also Appendix~\ref{sec:n1decay})
\begin{align}
 \Gamma(\no\to\gamma\gld) 
 =\frac{|C^{\gamma\tilde\chi_1}|^2\,m^5_{\no}}{48\pi\Mpl^2m^2_{3/2}}, 
\label{decay}
\end{align}
and $\Gamma(\no\to\gamma\gld)=0.21$ GeV for the bino-like neutralino
with $m_{\no}=300$ GeV and $m_{3/2}=10^{-13}$ GeV.
An irreducible SM background for the signal of mono-photon plus missing 
energy comes from $e^+e^-\to\gamma\nu\bar\nu$. 
In addition to the minimal cuts for the detection of photons 
\begin{align}
 E_{\gamma}>0.03\,\sqrt{s},\quad |\eta_{\gamma}|<2,
\label{mincut}
\end{align}
we impose the $Z$-peak cut
\begin{align}
 E_{\gamma}<\frac{s-m_Z^2}{2\sqrt{s}}-5\Gamma_Z,
\label{zcut}
\end{align}
which can remove the contributions from 
$e^+e^-\to\gamma Z\to\gamma\nu\bar\nu$. 
The most significant background coming from the $t$-channel
$W$-exchange process can be reduced by using polarized $e^{\pm}$ beams. 

In Table~\ref{table:photon}, the selection efficiencies for the signal
and background processes with different polarizations%
\footnote{$|P_{e^-}|>0.8$ and $|P_{e^+}|>0.5$ are designed at the
International Linear Collider (ILC)~\cite{:2007sg}.}
are presented,
where the above kinematical cuts, \eqref{mincut} and \eqref{zcut}, are
taken into account. The cross sections both for the signal and
background are calculated by {\tt MG/MEv4}~\cite{Alwall:2007st}
supporting gravitino
interactions~\cite{Hagiwara:2010pi,Mawatari:2011jy}. 
Here, we assume the branching ratio of the neutralino decay is unity,  
$B(\no\to\gamma\gld)=1$, although other decay modes can be significant
in some parameter space~\cite{Ambrosanio:1999iu}. 
Since the cross section with $e^{\pm}$ beam polarizations
$P_{e^{\pm}}$ $(|P_{e^{\pm}}|\leq 1)$ is given by
\begin{align}
 \sigma(P_{e^-},P_{e^+}) &= 2\sum_{\lam}
 \Big(\frac{1+P_{e^-}\lam}{2}\Big)\Big(\frac{1-P_{e^+}\lam}{2}\Big)\,
 \sig_{\lam},
\end{align}
the signal cross section for the bino-like neutralino can be
enhanced, while the background can be reduced quite effectively,
by using a positively polarized $e^-$ beam ($P_{e^-}>0$) and a 
negatively polarized $e^+$ beam ($P_{e^+}<0$).
It must be noted again that the signal cross section is inversely
proportional to the gravitino mass squared. 

\begin{figure}
 \epsfig{file=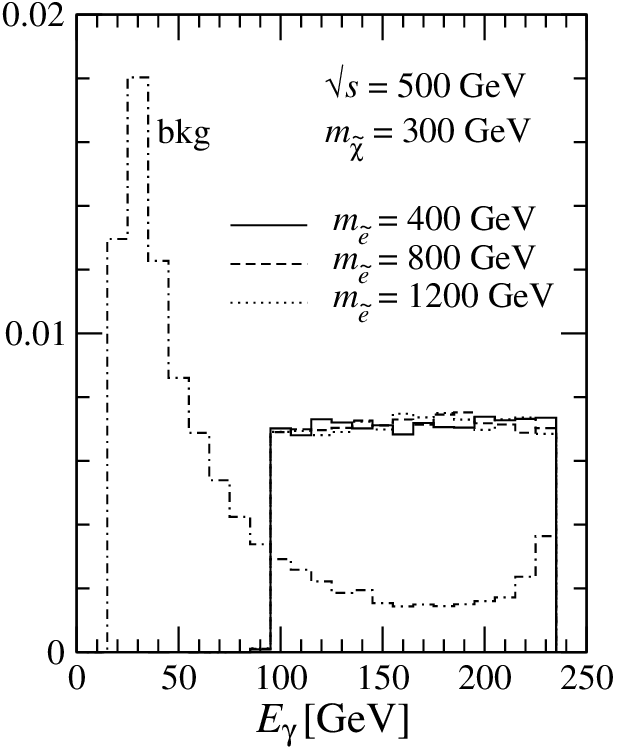,width=0.495\columnwidth,clip}
 \epsfig{file=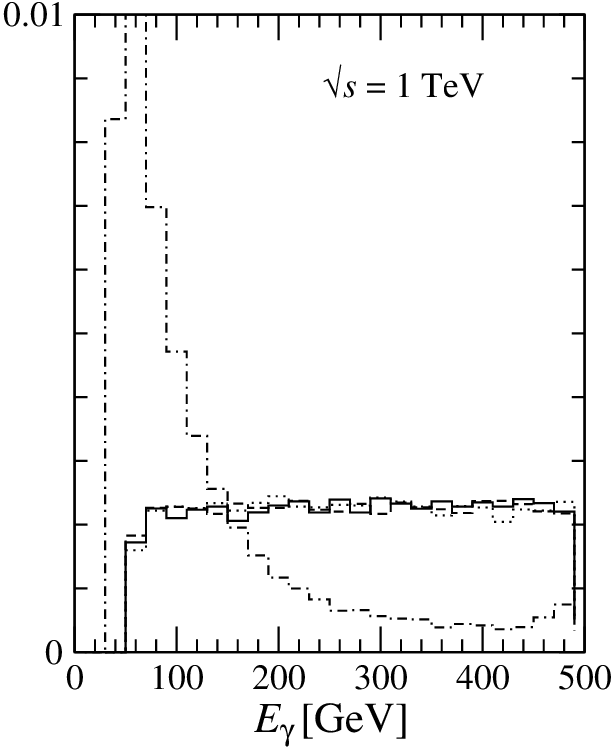,width=0.495\columnwidth,clip}
 \caption
 {Normalized energy distributions of the photon for
 $e^+e^-\to\no\gld\to\gamma\gld\gld$ at $\sqrt{s}=500$ GeV (left) and 1
 TeV (right), where $m_{\se_{\pm}}=400$ (solid), 800
 (dashed) and 1200 (dotted) GeV with $m_{\no}=300$ GeV are considered.
 The kinematical cuts
 in \eqref{mincut} and \eqref{zcut} and the beam polarizations 
 $(P_{e^-},P_{e^+})=(0.9,-0.6)$ are taken into account.
 Those of the SM background are also shown by dot-dashed lines.}
\label{fig:ea}
\end{figure} 

Figure~\ref{fig:ea} shows normalized energy distributions of the photon
for the signal and the SM background, corresponding to 20,000 events
each, at $\sqrt{s}=500$~GeV (left) and 1~TeV (right), where the
selectron mass of 400, 800 and 1200 GeV with the 300~GeV neutralino mass
are considered. The kinematical cuts in \eqref{mincut} and \eqref{zcut}
and the beam polarizations $(P_{e^-},P_{e^+})=(0.9,-0.6)$ are taken into
account. 
The signal distributions are flat, independent of the selectron mass,%
\footnote{We point out that the photonic energy distributions of Fig.~5 in 
 Ref.~\cite{Lopez:1996ey} should be flat and not depend on the selectron
 mass.}
and restricted as 
\begin{align}
 \frac{m_{\no}^2}{2\sqrt{s}}<E_{\gamma}<\frac{\sqrt{s}}{2},
\end{align}
where the lower edge can determine the mass of the neutralino. 
It should be stressed again that the photon distribution is simply given
by kinematical effects of the decaying neutralino.
The background is mostly distributed in the low-energy region, and
hence a further cut on the photon energy would be useful to enhance the
signal over background. 

\begin{figure}
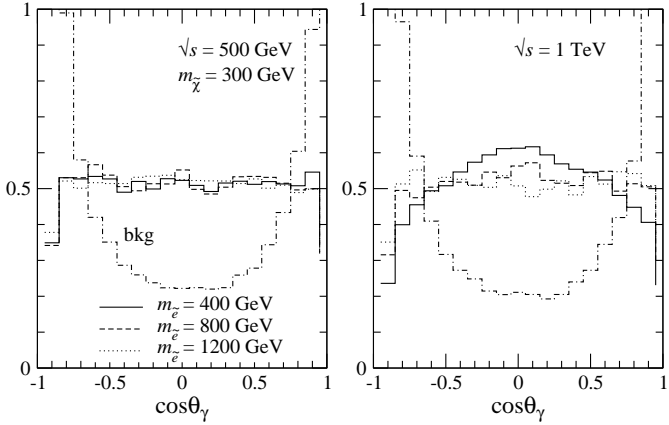

\centering 
 \epsfig{file=cosa_l.eps,width=0.49\columnwidth,clip}
 \epsfig{file=cosa_h.eps,width=0.49\columnwidth,clip}
 \caption
 {Normalized angular distributions of the photon in the laboratory frame
 for $e^+e^-\to\no\gld\to\gamma\gld\gld$. 
 The detail is the same as Fig.~\ref{fig:ea}.
 }
\label{fig:cosa}
\end{figure} 

Finally, we present the angular dependence of the photon in the
laboratory frame in Fig.~\ref{fig:cosa}. The original angular
distributions of the neutralino in Fig.~\ref{fig:cos} are flattened for
the case of a 500~GeV collider since the neutralino decays isotropically
in its rest frame and the boost effect is small.
On the other hand, the angular distributions still survive for the case 
of a 1~TeV collider. This indicates a possibility to examine the mass of
the $t,u$-channel selectrons when the decaying neutralino has a large
momentum. We note that a kinematical cut on the forward and backward
regions would also help to reduce the background.

\section{Selectron-gravitino production in $e^-\gamma$ collisions}
\label{sec:senlsp}

In this section, we repeat a study as in Sect.~\ref{sec:n1nlsp} for a
scenario of a selectron NLSP with a gravitino LSP. We consider
associated gravitino productions with a
selectron, especially a right-handed selectron, in $e^-\gamma$
collisions with the prompt selectron decay into an electron and a
gravitino, 
\begin{align}
 e^-\gamma\to\se^-_R\gld\to e^-\gld\gld,
\label{process2}
\end{align}
leading to a mono-electron plus missing-energy signal.

\subsection{Helicity amplitudes}\label{sec:amp2}

\begin{figure}
 \centering 
 \epsfig{file=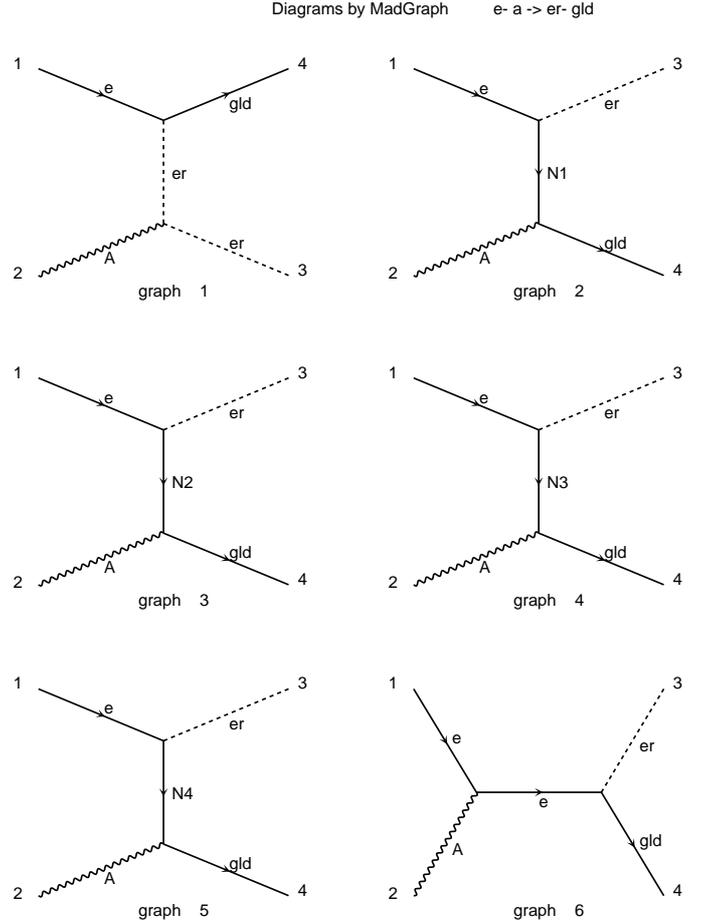,width=1\columnwidth,clip}
 \caption
 {Feynman diagrams for the $\se_R^-$-$\gld$ production in
 $e^-\gamma$ collisions, generated by 
 {\tt MadGraph}~\cite{Mawatari:2011jy}.}
\label{fig:diagram_se}
\end{figure}

Here we present the helicity amplitudes explicitly for the production
process: 
\begin{align}
 e^-\Big(p_1,\frac{\lam_1}{2}\Big)+\gamma(p_2,\lam_2)\to
 \se^-_R(p_3)+\gld\Big(p_4,\frac{\lam_4}{2}\Big).
\end{align}
The helicity amplitudes for the process are expressed as sums of $s$-,
$t$-, and $u$-channel amplitudes 
\begin{align}
 \M_{\lam_1\lam_2,\lam_4}=\M^s+\sum_{i=1}^4{\M^{t_i}}+\M^u,
\end{align}
corresponding to the Feynman graph 6, (2+3+4+5), and 1, respectively, in
Fig.~\ref{fig:diagram_se}. 
They are given in the four-spinor basis by 
\begin{subequations}
\begin{align}
 i\M^{s}_{\lam_1\lam_2,\lam_4}&=
  \frac{-e\,m^2_{\se_{\lam_1}}}{\sqrt{3}\,\Mpl m_{3/2}}\frac{1}{s}\,
   \epsilon_{\mu}(p_2,\lam_2)\nn\\
  &\times\bar{u}(p_4,\lam_4)
  (\slashed{p}_1+\slashed{p}_2)\gamma^{\mu}u(p_1,\lam_1),\\
 i\M^{t_i}_{\lam_1\lam_2,\lam_4}&=
  \frac{e\,m_{\tilde\chi^0_i}
        C^{\gamma\tilde\chi_i}_{}C^{\se\tilde\chi_i}_{\lam_1}}
       {2\sqrt{3}\,\Mpl m_{3/2}}
  \frac{1}{t-m^2_{\tilde\chi^0_i}}\,{\epsilon}_{\mu}(p_2,\lam_2) \nn\\
  &\times\bar{u}(p_4,\lambda_4)[\slashed{p}_2,\gamma^{\mu}]
  (\slashed{p}_1-\slashed{p}_3+m_{\tilde\chi^0_i})u(p_1,\lam_1),\\
 i\M^{u}_{\lam_1\lam_2,\lam_4}&=
  \frac{-e\,m^2_{\se_{\lam_1}}}{\sqrt{3}\,\Mpl m_{3/2}}
  \frac{1}{u-m^2_{\se_{\lam_1}}}\,
  \epsilon_{\mu}(p_2,\lam_2) \nn\\
  &\times\bar{u}(p_4,\lam_4)u(p_1,\lam_1)\,(p_3+p_1-p_4)^{\mu},
\end{align}
\end{subequations}
with the couplings $C^{\gamma\tilde\chi_i}$ and
$C^{\se\tilde\chi_i}_{\pm}$ defined in~\eqref{couplings}. 
Only the $\lam_1=+1$ case contributes to the $\se_R$ (or $\se_+$ in our
notation) production in the final state, which is relevant in our
following analyses. On the other hand, the $\lam_1=-1$ case gives
nonzero amplitudes only for the $\se_L$ production.

\begin{table*}
\centering
\begin{tabular}{cc|rcccccc}
 $\lam_1\lam_2$ &$ \lambda_4$& & 
  & $\hat{\mathcal{M}}^{s}$ & $\hat{\mathcal{M}}^{t} $ 
  & $\hat{\mathcal{M}}^{u}$ \\
 \hline
 $++$ & $-$ & $2\sin{\frac{\theta}{2}}$&$\big[$&
  $\frac{m^2_{\tilde{e}}}{s}$&
  $-\sum_{i}{C^{\gamma\tilde\chi_i}C^{\se\tilde\chi_i}_+
   \frac{m_{\tilde{\chi}_i}^2}{t-m^2_{\tilde{\chi}_i}}}$&
  $+\frac{m^2_{\tilde{e}}}{u-m^2_{\tilde{e}}}\beta
   \frac{1+\cos\theta}{2}$ & $\big]$ \\
 $+-$ & $+$ & $(1-\cos{\theta})\cos{\frac{\theta}{2}}$ &$\big[$& &
  $-\sum_{i}{C^{\gamma\tilde\chi_i}C^{\se\tilde\chi_i}_+
   \frac{\sqrt{s}m_{\tilde{\chi}_i}}{t-m^2_{\tilde{\chi}_i}}}\beta$
  && $\big]$ \\
 $+-$ & $-$ & $-(1+\cos{\theta})\sin{\frac{\theta}{2}}$&$\big[$& &&
  $\frac{m^2_{\tilde{e}}}{u-m^2_{\tilde{e}}}\beta$&$\big]$
\end{tabular}
 \caption
 {The reduced helicity amplitudes $\hat\M_{\lam_1\lam_2,\lam_4}$ for
 $e^-_{\lam_1}\gamma_{\lam_2}^{}\to\se_R^-\gld_{\lam_4}^{}$. }
\label{table:n2}
\end{table*}

We define the reduced helicity amplitudes, $\hat{\M}$, as 
\begin{align}
 i\M_{\lam_1\lam_2,\lam_4}=\frac{-e}{\sqrt{6}\,\Mpl
 m_{3/2}}\sqrt{\beta}\,s\,\hat{\M}_{\lam_1\lam_2,\lam_4},
\label{equation:reducedea}
\end{align}
and these are shown in Table~\ref{table:n2}. 
Similar to \eqref{equation_momenta_1}, the four-momenta
and helicities of the external particles are defined in the $e\gamma$ CM
frame with $\beta=1-m_{\se_R}^2/s$. 
The following features of the amplitudes are worth noting:
\begin{enumerate}
\item The amplitude $\M_{++,+}$ is zero since the coupling structures do
      not allow this helicity combination for a massless
      goldstino. \\[-2mm] 
\item The overall angular dependence is dictated by $J=1/2$ or $J=1$ $d$
      functions as
\begin{align}
 \M_{\lam_1\lam_2,\lam_4}\propto 
 d^{|\lam_1/2-\lam_2|}_{\lam_1/2-\lam_2,-\lam_4/2}(\theta).
\end{align}
\item The amplitude $\M^t$ depends on the mass of the propagating
      neutralinos; as the neutralino mass increases, $\M^t_{++,-}$
      becomes larger, while $\M^t_{+-,+}$ becomes smaller.  
      On the other hand, the $\M^s$ and $\M^u$ do not depend on their
      mass but on the selectron mass. \\[-2mm] 
\item The right-handed selectron can couple only to the bino component
      of neutralinos $\tilde\chi^0_i$, i.e. $U_{1i}$ in the
      $\tilde\chi^0_i$-$e$-$\se_R$ coupling
      in~\eqref{couplings}. Therefore, e.g. for the bino-like lightest
      neutralino case, only the $\no$-exchange amplitude is nonzero
      among the four neutralino amplitudes. \\[-2mm]
\item In the threshold region, similar to the $e^+e^-\to\no\gld$
      process, the amplitudes are proportional to $\beta^{3/2}$, which
      gives rise to the strong suppression on the production cross
      section. \\[-2mm] 
\item In the high-energy limit, the amplitude $\M_{+-,+}$ becomes
      dominant, and hence the cross section depends on the neutralino
      mass but not on the produced selectron mass. 
\end{enumerate}

We note that our helicity amplitudes in Table~\ref{table:n2} agree with
Eqs.~(4) and (5) in~\cite{Gopalakrishna:2001cm},%
\footnote{Except the sign in the parentheses of the first term in
 Eq.~(4) in~\cite{Gopalakrishna:2001cm}.}
where the heavy neutralino mass limit is assumed.
The helicity-summed amplitude squared also agrees with Eq.~(7)
of~\cite{Klasen:2006kb} for the squark associated process
$qg\to\tilde q\gld$ after substitutions for the masses and the couplings
as in~\eqref{substitute} and the exchange of $t\leftrightarrow u$.
Moreover, we checked our amplitudes for each helicity combination
numerically by the gravitino/goldstino code in 
{\tt MG/ME}~\cite{Hagiwara:2010pi,Mawatari:2011jy}.

\subsection{Cross sections and kinematical distributions}

\begin{figure}[b]
 \epsfig{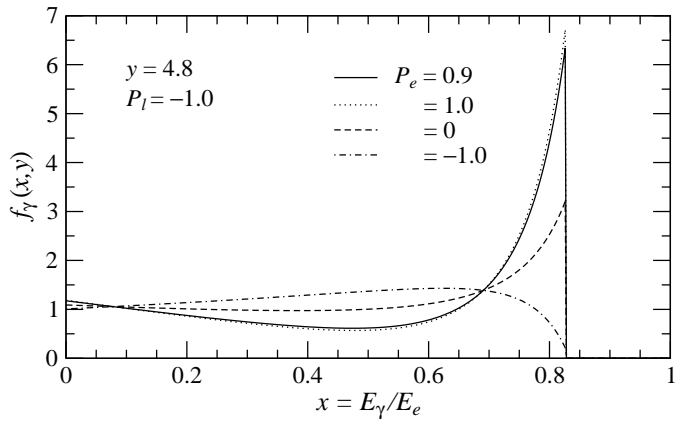}
 \caption
 {The distribution functions of Compton back-scattered photons in
 \eqref{pdf} for different electron beam polarizations.}
\label{fig:pdf}
\end{figure} 

In practice, a high-energy photon beam is provided by the backward
Compton scattering of laser photons on a high-energy 
electron beam~\cite{Ginzburg:1981ik,Ginzburg:1982yr}, as an option of a
future linear collider~\cite{:2007sg}. Let us introduce the photon
luminosity function~\cite{Ginzburg:1981ik,Badelek:2001xb} 
\begin{align}
 f_{\gamma}(x,y)=\frac{1}{N(y)}&\Big[\frac{1}{1-x}+1-x-4r(1-r)\nn \\
 &+P_eP_{l}\,ry(1-2r)(2-x)\Big],
\label{pdf}
\end{align}
where $x=E_{\gamma}/E_{e}$ is the energy ratio of the scattered photon
and the electron beam, $y$ is a parameter controlled by the laser
energy,%
\footnote{$y=4E_eE_l/m_e^2$ in the zero angle limit of the Compton
scattering.}  
$r=x/(1-x)y$, and $P_e$ ($P_{l}$) is the incident electron beam 
(laser photon) polarization. The integral $\int f_{\gamma}(x,y)\,dx$ is
normalized to unity by 
\begin{multline}
 N(y)=\Big(1-\frac{4}{y}-\frac{8}{y^2}\Big)\ln(1+y)+\frac{1}{2}
  +\frac{8}{y}-\frac{1}{2(1+y)^2} \\
  +P_{e}P_{l}\Big[\Big(1+\frac{2}{y}\Big)\ln(1+y)
   -\frac{5}{2}+\frac{1}{1+y}-\frac{1}{2(1+y)^2}\Big].
\end{multline}
Figure~\ref{fig:pdf} shows the luminosity function for different
electron beam polarizations, with $y=4.8$ and $P_{l}=-1.0$ as an optimal
parameter choice~\cite{Badelek:2001xb}. The maximal energy fraction is
fixed by $x_{\rm max}=y/(1+y)$, e.g. $x_{\rm max}\sim 0.83$ for
$y=4.8$. The distribution with highly polarized electron and laser beams 
($P_eP_l\sim -1$) has a strong peak at the high-energy endpoint.%
\footnote{Although the scattered photons are polarized when $P_e\ne0$
 or $P_l\ne0$~\cite{Ginzburg:1982yr,Badelek:2001xb}, we average the two
 opposite polarized modes, $(P_e>0,P_l<0)$ and $(P_e<0,P_l>0)$, so that
 we consider $f_{\gamma}(x,y)$ as the unpolarized distribution function
 in the following analyses.} 

The full cross section at an $e\gamma$ collider is calculated by
convoluting the $e\gamma$ cross section ($\sigma^{e\gamma}$) with the
photon distribution function of Eq.~\eqref{pdf} as 
\begin{align}
 \sig({s}_{ee})=\int_{x_{\rm min}}^{x_{\rm max}} 
  f_{\gamma}(x,y)\,\sig^{e\gamma}(s)\,dx
\label{totalx}
\end{align}
with $x_{\rm min}=m_{\se}^2/s_{ee}$ and $s=xs_{ee}$, where
$\sqrt{s_{ee}}$ is the original $e^-e^-$ CM energy. The spin-summed and
averaged $e\gamma$ cross section is obtained by  
$\sigma^{e\gamma}=(\sigma^{e\gamma}_++\sigma^{e\gamma}_-)/2$ with the
photon-helicity ($\lam_2$) dependent cross section
\begin{align}
 d\sigma^{e\gamma}_{\lam_2} &= \frac{1}{2s}\frac{1}{2}\sum_{\lam_{1,4}}
 |\M_{\lam_1\lam_2,\lam_4}|^2d\Phi_2.
\end{align}

\begin{figure}
 \epsfig{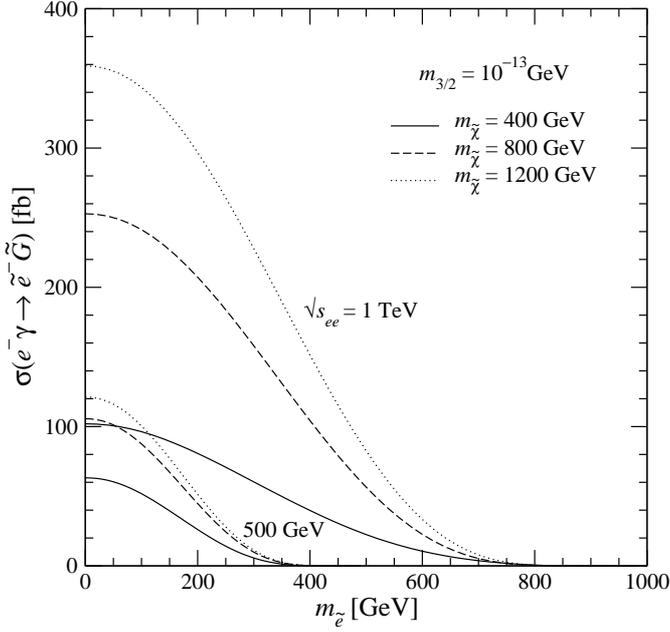}
 \caption
 {Total cross sections of associated gravitino productions with a 
 right-handed selectron in $e^-\gamma$ collisions,
 $e^-\gamma\to\se_R^-\gld$, at $\sqrt{s_{ee}}=500$ GeV and 1 TeV for
 $m_{3/2}=10^{-13}$ GeV as a function of the selectron mass. The
 neutralino mass is fixed at 400 (solid), 800 (dashed) and 1200 (dotted)
 GeV, respectively.}
\label{fig:xsec_mer}
\end{figure} 

Figure~\ref{fig:xsec_mer} shows total cross sections of the associated
gravitino productions with a right-handed selectron in $e^-\gamma$ 
collisions as a function of the selectron mass, where the CM energy,
$\sqrt{s_{ee}}$, of the $e^-e^-$ system is fixed at 500~GeV and 1~TeV.
The parameters for the photon luminosity function in \eqref{pdf} are
taken to be $y=4.8$ and $P_eP_l=-0.9$. For simplicity, we
assume a bino-like lightest neutralino so that only the
$\no$-exchange amplitude is taken into account for the $t$-channel
amplitude; see comment 4 in Sect.~\ref{sec:amp2}.
We note that $\sigma\propto 1/m_{3/2}^2$ and we take
$m_{3/2}=10^{-13}$~GeV in our study.  
When the selectron mass is close to the collider energy, the cross sections
are strongly suppressed due to $\sig^{e\gamma}\propto\beta^4$ as
mentioned in comment 5 in Sect.~\ref{sec:amp2}, similar
to the $e^+e^-\to\no\gld$ process in Fig.~\ref{fig:xsec_mn1}. 
In addition, the production cross section is nonzero only when
$m_{\se_R}<\sqrt{x_{\rm max}s_{ee}}$. It should be stressed here that
the cross section is quite sensitive to the mass of the $t$-channel
intermediate neutralinos, even if the collider energy cannot reach
their mass.

\begin{figure}
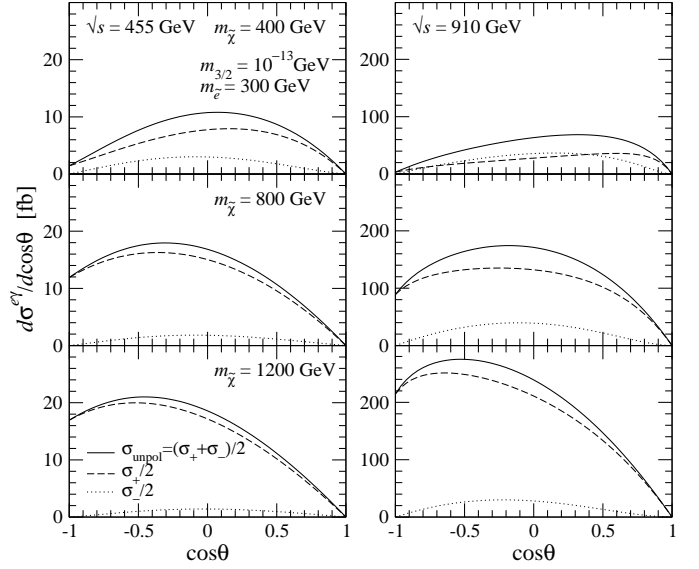

 \epsfig{file=coser_l.eps,height=7.45cm,clip}
 \epsfig{file=coser_h.eps,height=7.45cm,clip}
 \caption
 {Angular distributions of the selectron in $e^-\gamma\to\se_R^-\gld$ at
 $\sqrt{s}=455$ (left) and 910 (right) GeV in the $e\gamma$ CM frame,
 with $m_{3/2}=10^{-13}$ GeV and $m_{\se_R}=300$ GeV. The neutralino
 mass is taken to be 400 (top), 800 (middle) and 1200 (bottom) GeV,
 respectively. The photon-helicity dependent cross sections
 $d\sig^{e\gamma}_{\lam_2}$ are also shown by dashed lines ($\lam_2=+1$)
 and dotted lines ($\lam_2=-1$).}  
\label{fig:coser}
\end{figure} 

Before we consider the selectron decay, let us look in detail at the
angular distribution of the produced selectron in the $e\gamma$ system 
since the scalar decay $\se_R\to e\gld$ is isotropic and hence the
electron distribution is given by purely kinematical effects of the
decaying selectron and the boost from the $e^-\gamma$ CM frame to the
$e^-e^-$ laboratory frame.  
In Fig.~\ref{fig:coser}, the $\cos\theta$ distributions of the selectron
in $e^-\gamma\to\se_R^-\gld$ for $m_{3/2}=10^{-13}$~GeV and
$m_{\se_R}=300$~GeV are shown, where the scattering angle $\theta$ is
defined from the momentum direction of the incident electron in the
$e\gamma$ CM frame. 
We fix the $e\gamma$ CM energy as $\sqrt{s}=\sqrt{x_{\rm max}s_{ee}}$,
where the photon luminosity sharply peaks for $P_eP_l\sim-1$ (see
Fig.~\ref{fig:pdf}), i.e. 455~GeV for a 500~GeV collider (left) and
910~GeV for a 1~TeV collider (right). 
One can find that not only the total cross section as shown in
Fig.~\ref{fig:xsec_mer} but also the angular distribution is quite
sensitive to the mass of the $t$-channel intermediate neutralinos. As
the neutralino mass is increasing, the cross section with $\lam_2=+1$
becomes larger and the peak is shifted to the backward since the
$t$-channel amplitude becomes more important and its intrinsic
$\sin\frac{\theta}{2}$ angular dependence is revealed as
$1/(t-m_{\tilde\chi}^2)$ goes to $1/m_{\tilde\chi}^2$.  
On the other hand, the cross section with $\lam_2=-1$ becomes smaller.
As easily seen in Table~\ref{table:n2}, the productions to the forward
region ($\cos\theta=1$) are forbidden for all helicity combinations
because of the angular momentum conservation, while the productions to
the backward region ($\cos\theta=-1$) are allowed only for the
$\lam_2=+1$ case. 

Let us now turn to the simulations for the signal of single-electron
plus missing energy in the $e^-e^-$ laboratory frame. The partial decay
rate of the selectron decay into an electron and a gravitino is given by
\begin{align}
 \Gamma(\se_R\to e\gld) 
 =\frac{m^5_{\se_R}}{48\pi\Mpl^2m^2_{3/2}}, 
\end{align}
and $\Gamma(\se_R\to e\gld)=0.27$~GeV with $m_{\se_R}=300$~GeV and
$m_{3/2}=10^{-13}$~GeV. An irreducible SM background for the event of
mono-electron plus missing energy comes from 
$e^-\gamma\to e^-\nu\bar\nu$.  
In addition to the minimal cuts for the detection of electrons 
\begin{align}
 E_{e}>0.03\,\sqrt{s},\quad |\eta_{e}|<2,
\label{mincut2}
\end{align}
we impose the $Z$-peak cut
\begin{align}
 M_{\rm miss}>100\ {\rm GeV},
\label{zcut2}
\end{align}
which can remove the contributions from 
$e^-\gamma\to e^- Z\to e^-\nu\bar\nu$. The main background
contribution coming from the $W$-exchange can be reduced by using a
polarized electron beam.  

\begin{table}
\centering
\begin{tabular}{lr|rrr}
 $\qquad\sigma$ [fb] & & $P_{e^-}=$ & $0$ & $0.9$ \\
 \hline
 $\sqrt{s_{ee}}=500$ GeV & $m_{\tilde\chi}=400$ GeV & & 5 & 9 \\
 & 800 GeV & & 9 & 16  \\
 & 1200 GeV & & 10 & 18  \\
 & SM background & & 2594 & 284 \\
 \hline
 $\sqrt{s_{ee}}=1$ TeV & $m_{\tilde\chi}=400$ GeV & & 58 & 110 \\ 
 & 800 GeV & & 152 & 289 \\
 & 1200 GeV & & 220 & 416 \\
 & SM background & & 2796 & 290 
\end{tabular}
 \caption
 {Cross sections in fb unit for the signal, 
 $e^-\gamma\to\se_R^-\gld\to e^-\gld\gld$, 
 and the SM background, $e^-\gamma\to e^-\nu\bar\nu$, at
 $\sqrt{s_{ee}}=500$~GeV and 1~TeV without and with the electron beam
 polarization $P_{e^-}=0.9$.
 We take  $m_{3/2}=10^{-13}$~GeV, $m_{\se_R}=300$~GeV, and
$B(\se_R\to e\gld)=1$. 
 The minimal cuts in \eqref{mincut2} and the $Z$-peak cut
 in \eqref{zcut2} are taken into account.}   
\label{table:electron}
\end{table}

In Table~\ref{table:electron}, the selection efficiencies
for the signal and background processes without and with the electron
beam polarization are presented, where the above two
kinematical cuts, \eqref{mincut2} and \eqref{zcut2}, are taken into
account and it is assumed that the branching ratio of the selectron
decay to an electron and a gravitino is unity,
$B(\se_R\to e\gld)=1$. 
The cross sections both for the signal and background are calculated by
{\tt MG/MEv4}~\cite{Alwall:2007st}
with gravitino
interactions~\cite{Hagiwara:2010pi,Mawatari:2011jy}, where we also
implemented the photon luminosity function of~\eqref{pdf}. 
By using a positively polarized electron beam of $P_{e^-}=0.9$, 
the signal is enhanced by a factor of 1.9 because the cross section with
$\lam_1=-1$ is zero, while the background can be reduced by a factor of 10. 
It must be noted again that the signal cross section is inversely
proportional to the gravitino mass squared. 

\begin{figure}
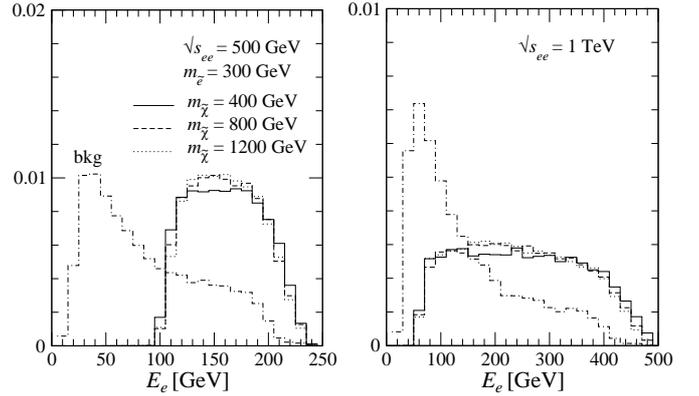

\centering 
 \epsfig{file=ee_l.eps,width=0.49\columnwidth,clip}
 \epsfig{file=ee_h.eps,width=0.49\columnwidth,clip}
 \caption
 {Normalized energy distributions of the electron for
 $e^-\gamma\to\se_R^-\gld\to e^-\gld\gld$ at $\sqrt{s_{ee}}=500$ GeV
 (left) and 1 TeV (right), where
 $m_{\tilde\chi}=400$ (solid), 800 (dashed) and 1200 (dotted) GeV with 
 $m_{\se_R}=300$ GeV are considered.
 The kinematical cuts in \eqref{mincut2} and
 \eqref{zcut2} and the electron beam polarization $P_{e^-}=0.9$ are taken
 into account. 
 Those of the SM background are also shown by dot-dashed lines.}
\label{fig:ee}
\end{figure} 

Figure~\ref{fig:ee} presents normalized energy distributions of the
electron for the signal and
the SM background, corresponding to 20,000 events each, at
$\sqrt{s_{ee}}=500$ GeV (left) and 1 TeV (right), where
the lightest neutralino mass of 400, 800 and 1200 GeV with the 300 GeV
selectron mass are considered.
The kinematical cuts in \eqref{mincut2} and \eqref{zcut2} and the
electron beam polarization $P_{e^-}=0.9$ are taken into account.
We notice again that the electron distribution is given by
two boost effects, along the momentum of the decaying selectron and 
along the beam axis. The momentum of the incident electron is chosen
to the $+z$ direction, and hence the produced electrons in the
$e\gamma$ CM frame are boosted to the forward direction.
Although the signal distributions no longer have either a flat shape or
a sharp edge due to the boost along the beam direction, the energy is
restricted as 
\begin{align}
 \frac{m_{\se_R}^2}{2\sqrt{s_{ee}}}<E_{\gamma}<\frac{\sqrt{s_{ee}}}{2}, 
\end{align}
where the lower edge can determine the selectron mass.
The energetic electrons tend to be suppressed since the original
selectron productions in the forward region
are not allowed. 
Moreover, the $z$-axis boost effect makes the distributions slightly
different for the different neutralino mass, which can be dictated by
the peak shift in Fig.~\ref{fig:coser}.
Since the background is located mostly in the low-energy region, 
a certain amount of the energy cut can help to enhance the signal over
the background.

\begin{figure}
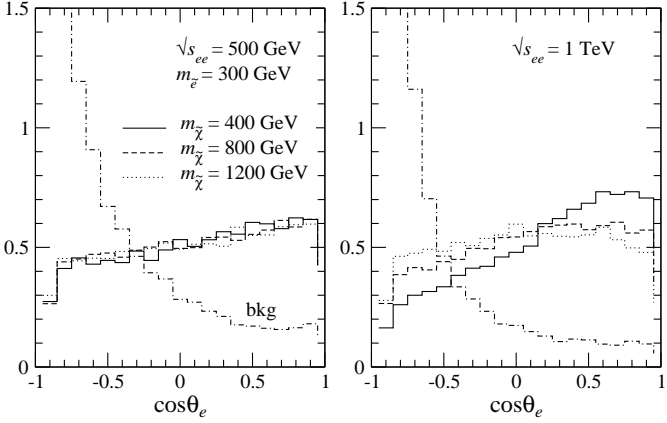

\centering 
 \epsfig{file=cose_l.eps,width=0.49\columnwidth,clip}
 \epsfig{file=cose_h.eps,width=0.49\columnwidth,clip}
 \caption{Normalized angular distributions of the electron in the
 $e^-e^-$ laboratory frame for $e^-\gamma\to\se_R^-\gld\to e^-\gld\gld$.
 The detail is the same as Fig.~\ref{fig:ee}.}
\label{fig:cose}
\end{figure} 

Finally, the angular distributions of the electron are shown in
Fig.~\ref{fig:cose}, where the angle $\theta_e$ is measured from the
direction of the electron beam, or the $+z$ direction,  in the $e^-e^-$
laboratory frame. 
The electrons tend to be produced more in the forward region 
($\cos\theta_e>0$) due to the system boost.
For the $\sqrt{s_{ee}}=500$ GeV case the original $m_{\tilde\chi}$
dependence in the angular distributions of
the selectron shown in Fig.~\ref{fig:coser} is no longer observed, while 
the dependence can be seen at $\sqrt{s_{ee}}=1$~TeV.
This indicates that we would be able to 
determine the mass of the $t$-channel neutralinos when the collider
energy is relatively higher than the selectron mass.
We note that the background can be reduced further by a
kinematical cut on the backward region. 

Before closing this section, we point out that selectron-neutralino
associated productions in neutralino LSP scenarios lead to the same
signal, or $e+\slashed E$, and have been studied 
intensively~\cite{Cuypers:1992mb,Choudhury:1994vi,Kiers:1996ux,Barger:1997qu}.
Since the LSP mass is quite different between the two models, 
${\cal O}$(eV) for the $\gld$ LSP
and ${\cal O}$(100 GeV) for the $\no$ LSP, the distributions of the
final electron are distinctive and could provide a hint of SUSY breaking 
mechanism.

\section{Summary}\label{sec:summary}

Associated gravitino productions with a SUSY particle can be observed in
current and future collider experiments if the gravitino is very light. 
In this paper, we restudied the two associated-gravitino-production
processes for a future $e^+e^-/e^-\gamma$ collider  
by using the gravitino implemented {\tt MadGraph/MadEvent}. 

First, we studied gravitino productions in association with a
neutralino which promptly decays into a photon and a gravitino at
an $e^+e^-$ collider, $e^+e^-\to\no\gld\to\gamma\gld\gld$.
By using the effective goldstino interaction Lagrangian
we explicitly presented the helicity amplitudes for the production
process, which give us deep understanding for the threshold behavior
and the SUSY-mass dependence of the production cross section and the
angular distributions.
We also examined selection efficiencies by kinematical cuts and beam
polarizations for the signal and SM background processes, and showed
that the energy and angular distributions of the photon in the final
state can explore the mass of the $t$-channel exchange selectrons as
well as the mass of the decaying neutralino. 

Second, we considered gravitino productions associated with a
selectron which subsequently decays into an electron and a gravitino at 
an $e\gamma$ collider, $e^-\gamma\to\se^-\gld\to e^-\gld\gld$.
We repeated the same analyses as in the first process;
we presented the explicit helicity amplitudes for the production
process, and discussed the mono-electron plus missing-energy signal,
including the energy spectrum of the 
backward-Compton scattered photons for incident photons. 
Similar to the $e^+e^-\to\no\gld$ process, we found that the production
cross section and the kinematical distributions of the electron in the
final state are quite sensitive to the mass of the $t$-channel
intermediate neutralinos as well as the mass of the decaying selectron.

We finally note that, throughout our study, we carefully checked our
calculations with the previous works both analytically and 
numerically, and pointed out a few disagreements. 

Before closing, we recall that all the helicity amplitudes we presented
are easily applicable to 
$q\bar q\to\go\gld$ and to
$qg\to\sq\gld$ subprocesses for hadron colliders.

\begin{acknowledgement}{\textit{Acknowledgements}}
We wish to thank Alberto Mariotti for valuable discussions and comments.
We also thank Fabio Maltoni and the members of the CP3, 
U. Catholique de Louvain for their warm hospitality, where part
 of this work has been done.
The work presented here has been in part supported by the Concerted
 Research action 
``Supersymmetric Models and their Signatures at the Large Hadron
 Collider'' 
of the Vrije Universiteit Brussel,
by the Belgian Federal Science Policy Office through the Interuniversity
 Attraction Pole IAP VI/11, 
and by the Grant-in-Aid for Scientific Research (No. 20340064) from the
 Japan Society for the Promotion of Science.  
\end{acknowledgement}

\appendix
\section{Effective goldstino interaction Lagrangian}
\label{sec:lagrangian}

We briefly present the relevant terms of the interaction Lagrangian
for our study. 
The effective goldstino interaction Lagrangian among goldstino, electron 
and selectron, $\gld$-$e$-$\se_{\pm}$, and among goldstino, neutralino
and photon/$Z$-boson, $\gld$-$\tilde\chi^0_i$-$V(=\gamma/Z)$, in
non-derivative form is
\begin{multline}
 {\cal L}_{\gld}
 =\mp\frac{im_{\se_{\pm}}^2}{\sqrt{3}\,\Mpl\,m_{3/2}}
   \big[ \bar{\psi}_{\gld}P_{\pm}\psi_e^{}\,\phi_{\se_{\pm}}^*
        -\bar{\psi}_e^{}P_{\mp}\psi_{\gld}\,\phi_{\se_{\pm}}^{} 
   \big] \\
 -\frac{C^{V\tilde{\chi}_i}m_{\tilde{\chi}^0_i}}
       {4\sqrt{6}\,\Mpl\,m_{3/2}}
  \bar{\psi}_{\gld}[\gamma^{\mu},\gamma^{\nu}]\psi_{\tilde{\chi}^0_i}
  (\partial_{\mu}V_{\nu}-\partial_{\nu}V_{\mu}), 
\label{L_int}
\end{multline}
where $\se_{\pm}$ denotes the right-/left-handed selectron, 
$P_{\pm}=\frac{1}{2}(1\pm\gamma_5)$ is the chiral-projection operator, 
and the coupling $C^{V\tilde{\chi}_i}$ is defined in~\eqref{couplings};
see more details in~\cite{Mawatari:2011jy}.
All other relevant terms are 
\begin{align}
 {\cal L}_{eeV}&=e\,\bar\psi_e^{}\gamma^{\mu}
   [A_{\mu}-(g_{+}P_{+}+g_{-}P_{-})Z_{\mu}]\psi_e^{}, \\
 {\cal L}_{{\tilde{\chi}}_i^0e\se}&=\pm\sqrt{2}\,e\,
  C^{\se\tilde{\chi}_i}_{\pm}
  \big[ \bar{\psi}_{\tilde\chi^0_{i}}P_{\pm}\psi_e^{}\,\phi_{\se_{\pm}}^*
       +\bar{\psi}_e^{}P_{\mp}\psi_{\tilde\chi^0_i}\,
        \phi_{\se_{\pm}}^{}\big], \\
 {\cal L}_{\se\se\gamma}&=ie\,
  \phi_{\se_{\pm}}^*\overleftrightarrow{\partial^{\mu}}
  \phi_{\se_{\pm}}^{}A_{\mu},
\end{align}
where $g_{\pm}$ and $C^{\se\tilde{\chi}_i}_{\pm}$ are defined 
in~\eqref{zcouplings} and~\eqref{couplings}, respectively.

\section{Helicity amplitudes for $\no\to\gamma\gld$}
\label{sec:n1decay}

We show helicity amplitudes for the neutralino decay into a photon and a
gravitino,  
\begin{align}
 \no\Big(p_1,\frac{\lam_1}{2}\Big)\to
 \gamma(p_2,\lam_2)+\gld\Big(p_3,\frac{\lam_3}{2}\Big).
\end{align}
The partial decay rate in the neutralino rest frame is given by
\begin{align}
 \Gamma&=\frac{1}{2m_{\no}}\frac{1}{2}\int\sum_{\lam_{1,2,3}}|
         \M_{\lam_1,\lam_2\lam_3}|^2d\Phi_2,
\end{align}
and the helicity amplitudes are calculated as
\begin{align}
 \M_{+,++}&=-\M_{-,--}=
  \frac{-C^{\gamma\tilde\chi_1}m^3_{\no}}{\sqrt{3}\,\Mpl m_{3/2}}
  \cos{\frac{\theta^*}{2}},\nn\\
 \M_{+,--}&=\M_{-,++}=
  \frac{-C^{\gamma\tilde\chi_1}m^3_{\no}}{\sqrt{3}\,\Mpl m_{3/2}}
  \sin{\frac{\theta^*}{2}},
\end{align}
with $C^{\gamma\tilde\chi_1}$ in \eqref{couplings} and the decay angle
$\theta^*$ defined from the quantization axis of the neutralino spin. 
The angular dependence is dictated by $J=1/2$ $d$ functions as 
\begin{align}
 \M_{\lam_1,\lam_2\lam_3}\propto 
 d^{1/2}_{\lam_1/2,\,\lam_2-\lam_3/2}(\theta^*).
\end{align}
Summing over the initial or final helicities for the amplitudes squared
gives the isotropic decay distribution in the rest frame, and one can
find the well-known decay rate in Eq.~\eqref{decay}.



\begin{thebibliography}{00}

\bibitem{Giudice:1998bp}
  See, e.g.,
  G.~F.~Giudice and R.~Rattazzi,
  Phys.\ Rept.\  {\bf 322} (1999) 419.

\bibitem{Ellis:1984kd}
  J.~R.~Ellis, K.~Enqvist, D.~V.~Nanopoulos,
  Phys.\ Lett.\  {\bf B147} (1984)  99;
%
  ibid.\ {\bf B151} (1985)  357.

\bibitem{Lopez:1992ni}
  J.~L.~Lopez, D.~V.~Nanopoulos and A.~Zichichi,
  Phys.\ Rev.\  D {\bf 49} (1994) 343
  [arXiv:hep-ph/9210280];
%
  Int.\ J.\ Mod.\ Phys.\  A {\bf 10} (1995) 4241
  [arXiv:hep-ph/9408345].

\bibitem{Gherghetta:2000qt}
  T.~Gherghetta and A.~Pomarol,
  Nucl.\ Phys.\  B {\bf 586} (2000) 141
  [arXiv:hep-ph/0003129];
%
  ibid. {\bf B602 } (2001)  3-22.
  [hep-ph/0012378].

\bibitem{Fayet:1986zc}
  P.~Fayet,
  Phys.\ Lett.\  B {\bf 175} (1986) 471.

\bibitem{Dicus:1990vm}
  D.~A.~Dicus, S.~Nandi and J.~Woodside,
  Phys.\ Lett.\  B {\bf 258} (1991) 231.

\bibitem{Lopez:1996gd}
  J.~L.~Lopez, D.~V.~Nanopoulos and A.~Zichichi,
  Phys.\ Rev.\ Lett.\  {\bf 77}, 5168 (1996)
  [arXiv:hep-ph/9609524].

\bibitem{Lopez:1996ey}
  J.~L.~Lopez, D.~V.~Nanopoulos and A.~Zichichi,
  Phys.\ Rev.\  D {\bf 55}, 5813 (1997)
  [arXiv:hep-ph/9611437].

\bibitem{Gopalakrishna:2001cm}
  S.~Gopalakrishna and J.~D.~Wells,
  Phys.\ Lett.\ B {\bf 518} (2001) 123
  [arXiv:hep-ph/0108006].

\bibitem{Dicus:1989gg}
  D.~A.~Dicus, S.~Nandi and J.~Woodside,
  Phys.\ Rev.\  D {\bf 41} (1990) 2347;
%
  D.~A.~Dicus and S.~Nandi,
  Phys.\ Rev.\  D {\bf 56} (1997) 4166
  [arXiv:hep-ph/9611312].

\bibitem{Kim:1997iwa}
  J.~Kim, J.~L.~Lopez, D.~V.~Nanopoulos, R.~Rangarajan and A.~Zichichi,
  Phys.\ Rev.\  D {\bf 57} (1998) 373
  [arXiv:hep-ph/9707331].

\bibitem{Klasen:2006kb}
  M.~Klasen and G.~Pignol,
  Phys.\ Rev.\  D {\bf 75} (2007) 115003
  [arXiv:hep-ph/0610160].

\bibitem{Abdallah:2003np}
  J.~Abdallah {\it et al.} [DELPHI Collaboration],
  Eur.\ Phys.\ J.\ C {\bf 38} (2005) 395
  [arXiv:hep-ex/0406019].

\bibitem{Acosta:2002eq}
  D.~E.~Acosta {\it et al.} [CDF Collaboration],
  Phys.\ Rev.\ Lett.\ {\bf 89} (2002) 281801
  [arXiv:hep-ex/0205057].

\bibitem{Affolder:2000ef}
  A.~A.~Affolder {\it et al.} [CDF Collaboration],
  Phys.\ Rev.\ Lett.\ {\bf 85} (2000) 1378
  [arXiv:hep-ex/0003026].

\bibitem{Brignole:1998me}
  A.~Brignole, F.~Feruglio, M.~L.~Mangano and F.~Zwirner,
  Nucl.\ Phys.\  B {\bf 526} (1998) 136
  [Erratum-ibid.\  B {\bf 582} (2000) 759]
  [arXiv:hep-ph/9801329].

\bibitem{Hagiwara:2010pi}
  K.~Hagiwara, K.~Mawatari and Y.~Takaesu,
  Eur.\ Phys.\ J.\ C {\bf 71} (2011) 1529
  [arXiv:1010.4255 [hep-ph]].

\bibitem{Hagiwara:1990dw}
  K.~Hagiwara, H.~Murayama and I.~Watanabe,
  Nucl.\ Phys.\  B {\bf 367} (1991) 257;
%
  H.~Murayama, I.~Watanabe and K.~Hagiwara,
  KEK-Report 91-11, 1992.

\bibitem{Stelzer:1994ta}
  T.~Stelzer and W.~F.~Long,
  Comput.\ Phys.\ Commun.\  {\bf 81} (1994) 357
  [arXiv:hep-ph/9401258];
%
  G.~C.~Cho, K.~Hagiwara, J.~Kanzaki, T.~Plehn, D.~Rainwater and T.~Stelzer,
  Phys.\ Rev.\  D {\bf 73} (2006) 054002
  [arXiv:hep-ph/0601063].

\bibitem{Maltoni:2002qb}
  F.~Maltoni and T.~Stelzer,
  JHEP {\bf 0302} (2003) 027
  [arXiv:hep-ph/0208156].

\bibitem{Alwall:2007st}
  J.~Alwall, P.~Demin, S.~de Visscher, R.~Frederix, M.~Herquet,
  F.~Maltoni, T.~Plehn, D.~Rainwaterd and T.~Stelzer,
  JHEP {\bf 0709} (2007) 028
  [arXiv:0706.2334 [hep-ph]];\\
  {\tt http://madgraph.phys.ucl.ac.be/}.

\bibitem{Mawatari:2011jy}
  K.~Mawatari and Y.~Takaesu,
  Eur.\ Phys.\ J.\  C {\bf 71} (2011) 1640
  [arXiv:1101.1289 [hep-ph]].

\bibitem{:2007sg}
  J.~Brau {\it et al.} [ILC Collaboration],
  arXiv:0712.1950 [physics.acc-ph].

\bibitem{Ginzburg:1981ik}
  I.~F.~Ginzburg, G.~L.~Kotkin, V.~G.~Serbo and V.~I.~Telnov,
  JETP Lett.\ {\bf 34} (1981) 491;
%
  I.~F.~Ginzburg, G.~L.~Kotkin, V.~G.~Serbo and V.~I.~Telnov,
  Nucl.\ Instrum.\ Meth.\  {\bf 205} (1983) 47.

\bibitem{Ginzburg:1982yr}
  I.~F.~Ginzburg, G.~L.~Kotkin, S.~L.~Panfil, V.~G.~Serbo and V.~I.~Telnov,
  Nucl.\ Instrum.\ Meth.\  A {\bf 219} (1984) 5.

\bibitem{Choi:2006mr}
  S.~Y.~Choi, K.~Hagiwara, H.~U.~Martyn, K.~Mawatari and P.~M.~Zerwas,
  Eur.\ Phys.\ J.\  C {\bf 51}, 753 (2007)
  [arXiv:hep-ph/0612301].

\bibitem{Ambrosanio:1999iu}
  S.~Ambrosanio and G.~A.~Blair,
  Eur.\ Phys.\ J.\  C {\bf 12} (2000) 287
  [arXiv:hep-ph/9905403].

\bibitem{Badelek:2001xb}
  B.~Badelek {\it et al.} [ECFA/DESY Photon Collider Working Group],
  Int.\ J.\ Mod.\ Phys.\ A {\bf 19} (2004) 5097
  [arXiv:hep-ex/0108012].

\bibitem{Cuypers:1992mb}
  F.~Cuypers, G.~J.~van Oldenborgh and R.~Ruckl,
  Nucl.\ Phys.\ B {\bf 383} (1992) 45
  [arXiv:hep-ph/9205209];
%
  H.~Konig and K.~A.~Peterson,
  Phys.\ Lett.\ B {\bf 294} (1992) 110
  [arXiv:hep-ph/9205201];
%
  T.~Kon and A.~Goto,
  Phys.\ Lett.\ B {\bf 295} (1992) 324.

\bibitem{Choudhury:1994vi}
  D.~Choudhury and F.~Cuypers,
  Nucl.\ Phys.\ B {\bf 451} (1995) 16
  [arXiv:hep-ph/9412245].

\bibitem{Kiers:1996ux}
  K.~Kiers, J.~N.~Ng and G.~h.~Wu,
  Phys.\ Lett.\ B {\bf 381} (1996) 177
  [arXiv:hep-ph/9604338].

\bibitem{Barger:1997qu}
  V.~D.~Barger, T.~Han and J.~Kelly,
  Phys.\ Lett.\ B {\bf 419} (1998) 233
  [arXiv:hep-ph/9709366].

\end{thebibliography}
\end{document}